\documentstyle[aps,epsf,axodraw]{revtex}  

\newcommand{\re}{\mbox{$\rm e$}}
\newcommand{\ri}{\mbox{$\rm i$}}

\newcommand{\bfm}[1]{\mbox{\boldmath$#1$}}
\newcommand{\ratio}[2]{\mbox{$#1\over#2$}}

\begin{document}        

\baselineskip 14pt
\title{
\hfill \raisebox{2ex}{\rm\small ISU-HET-99-2} \bigskip \\   
$\bfm{C\!P}$ Violation in  $\Omega^-$  Decays\thanks{Talk 
presented at DPF `99, Los Angeles, 5-9 January 1999.}   
}
\author{Jusak Tandean}
\address{Department of Physics and Astronomy, Iowa State University, 
Ames, IA 50011}
%
\maketitle              

\begin{abstract}        
We estimate the $C\!P$-violating rate asymmetry for the decay  
$\Omega^-\rightarrow\Xi\pi$. 
Within the standard model,  we find that it could be as large as  
$2\times 10^{-5}$.  
This is significantly larger than the corresponding rate asymmetries 
for other non-leptonic hyperon decays, which are typically less than 
$10^{-6}$.
\\ 
\end{abstract}   	

\bigskip 


The origin of  $C\!P$  violation remains a mystery in particle physics.  
So far,  $C\!P$-odd signals have been observed only in the kaon 
systems. 
To determine the source and nature of  $C\!P$  violation, it is 
necessary to observe it in several different processes.  
For a number of years theoretical efforts have been made to study it 
in the weak decays of hyperons belonging to the baryon 
octet~\cite{donpa,steger,heval}.   
Here, we present the results of a recent study exploring  $C\!P$  
violation in the nonleptonic decays of 
the  $\Omega^-$  hyperon~\cite{TanVal2}.

The $C\!P$-violating observable that we are considering is 
the rate-asymmetry for  $\,\Omega^-\rightarrow\Xi\pi\,$  decays. 
It is given by   
\begin{eqnarray}
\Delta \bigl( \Xi^0\pi^- \bigr)  \;=\;  
{ \Gamma \bigl( \Omega^-\rightarrow\Xi^0\pi^- \bigr) - 
 \Gamma \bigl( \overline{\Omega}{}^-\rightarrow\overline{\Xi}{}^0\pi^+ \bigr)  
 \over  
 \Gamma \bigl( \Omega^-\rightarrow\Xi^0\pi^- \bigr) + 
 \Gamma \bigl( \overline{\Omega}{}^-\rightarrow\overline{\Xi}{}^0\pi^+ \bigr) 
}   \;.  
\end{eqnarray}  
In order to evaluate this quantity in more detail, 
we parametrize the decay amplitude in the form
\begin{eqnarray}   \label{amplitude}     
\ri {\cal M}_{\Omega^-\rightarrow\Xi\pi}^{}  
\;=\;  
G_{\rm F}^{} m_{\pi}^2\; 
{\alpha_{\Omega^-\Xi}^{\rm (P)}\over \sqrt{2}\, f_{\!\pi}^{}}\,  
\bar{u}_\Xi^{}\, k_\mu^{}\, u_\Omega^\mu      \;,
\end{eqnarray}    
where  the  $u$'s  are baryon spinors,  $k$  is the outgoing 
four-momentum of the pion, $f_{\!\pi}^{}$  is the pion-decay 
constant, and  only the~dominant P-wave piece of the~amplitude 
is included.  
We will consider only the~P-wave because, experimentally, 
the~asymmetry parameter in these decays is small and consistent with 
zero~\cite{pdb}, indicating that they are dominated by a~P-wave.  
This amplitude has both  $\,|\Delta\bfm{I}|=1/2\,$  and 
$\,|\Delta\bfm{I}|=3/2\,$  components which are, in general, complex. 
We write the amplitudes for  
$\,\Omega^-\rightarrow\Xi^0\pi^-,\Xi^-\pi^0\,$  as 
\begin{eqnarray}   \label{isolabels}
\alpha_{\Omega^-\Xi^0}^{\rm (P)}  \;=\;   
\ratio{1}{\sqrt{3}} \left( 
\sqrt{2}\, \alpha^{(\Omega)}_1  
\re^{{\rm i}\delta_1^{} + {\rm i}\phi_1^{}}
\,-\,  \alpha^{(\Omega)}_3 \re^{{\rm i}\delta_3^{} + {\rm i}\phi_3^{}}
\right)   
\;, \hspace{3em}   
\alpha_{\Omega^-\Xi^-}^{\rm (P)}  \;=\;   
\ratio{1}{\sqrt{3}} \left( 
\alpha^{(\Omega)}_1 \re^{{\rm i}\delta_1^{} + {\rm i}\phi_1^{}} 
\,+\, \sqrt{2}\, \alpha^{(\Omega)}_3  
\re^{{\rm i}\delta_3^{} + {\rm i}\phi_3^{}}
\right)   \;,   
\end{eqnarray}  
where  $\alpha^{(\Omega)}_{1,3}$  are real quantities,  
strong-rescattering phases of the  $\Xi\pi$  system with  $\,J=3/2$,  
P-wave and  $\,I=1/2, 3/2\,$  quantum numbers are denoted by  
$\delta_{1}$, $\delta_{3}$,  respectively, and    
$C\!P$-violating weak phases are labeled  $\phi_{1}$,  $\phi_{3}$. 
The corresponding expressions for the antiparticle decay 
$\,\overline{\Omega}{}^-\rightarrow \overline{\Xi}\pi\,$  are 
obtained by changing the sign of the weak phases  $\phi_{1,3}^{}$  
in~(\ref{isolabels}). 
It follows that, to first order in the small ratio     
$\,\alpha_{3}^{(\Omega)}/\alpha_{1}^{(\Omega)},\,$   
\begin{eqnarray}
\Delta \bigl( \Xi^0\pi^- \bigr)  \;=\;  
\sqrt{2}\; {\alpha_{3}^{(\Omega)}\over \alpha_{1}^{(\Omega)}}\; 
\sin \bigl( \delta_3^{}-\delta_1^{} \bigr) \, 
\sin \bigl( \phi_3^{}-\phi_1^{}\bigr)   \;.   
\label{cpobs}  
\end{eqnarray}  
Similarly, 
$\, \Delta \bigl( \Xi^-\pi^0 \bigr) 
= -2 \Delta \bigl( \Xi^0\pi^- \bigr) .\,$

Lets evaluate the three factors in (\ref{cpobs})  one by one.   
Using the measured decay rates~\cite{pdb} and  ignoring all the phases, 
we can extract
$\, \alpha_{3}^{(\Omega)}/\alpha_{1}^{(\Omega)} 
=-0.07\pm 0.01\;$~\cite{TanVal1}.      
Final-state interactions enhance this value, but the enhancement 
is not significant for the values of the scattering phases that we 
estimate below.   
This ratio is higher than the corresponding ratios in other hyperon 
decays~\cite{etv},  which range from  $0.03$  to  $0.06$ in magnitude, 
and provides an enhancement factor for the $C\!P$-violating rate 
asymmetry in this mode.

We now turn to the $\Xi\pi$-scattering  phases,  $\delta_{1,3}$.  
There exists no experimental information on these phases, and so we 
will estimate them at leading order in heavy-baryon chiral 
perturbation theory. 
The lowest-order chiral Lagrangian for the strong interactions of 
the octet and decuplet baryons with the pseudoscalar 
octet-mesons~\cite{JenMan} generates the diagrams shown 
in Fig.~\ref{diagrams} for  $\,\Xi^0\pi^-\rightarrow\Xi^0\pi^-\,$   
and  $\,\Xi^-\pi^0\rightarrow\Xi^-\pi^0.\,$   
From the resulting scattering amplitudes, we can construct 
amplitudes for the  $\,I=1/2\,$  and  $\,I=3/2\,$  channels 
using the relations    
\begin{eqnarray}
\begin{array}{c}   \displaystyle   
{\cal M}_{I=1/2}^{}  \;=\;  
2{\cal M}_{\Xi^0\pi^-\rightarrow \Xi^0\pi^-}^{}  
\,-\,  {\cal M}_{\Xi^-\pi^0\rightarrow \Xi^-\pi^0}^{}   \;,  
\vspace{2ex} \\   \displaystyle
{\cal M}_{I=3/2}^{}  \;=\;  
-{\cal M}_{\Xi^0\pi^-\rightarrow \Xi^0\pi^-}^{}
\,+\,  2{\cal M}_{\Xi^-\pi^0\rightarrow \Xi^-\pi^0}^{}   \;,  
\end{array}
\end{eqnarray}
and project out the partial waves in the usual way.
Calculating the $\,J=3/2\,$  P-wave phases, and evaluating them 
at a center-of-mass energy equal to the $\Omega^-$ mass, 
yields~\cite{TanVal2}    
\begin{eqnarray}   
\delta_1^{}  \;=\;  -12.8^{\small\rm o}  
\;, \hspace{3em}   
\delta_3^{}  \;=\;  1.1^{\small\rm o}   \;.    
\end{eqnarray}  
The  $\,I=1/2\,$  P-wave phase is larger than other baryon-pion 
scattering phases.   
For instance, the P-wave  $\Lambda\pi$-scattering 
phase has been estimated to be  
$\,\delta_{\rm P}^{}\approx -1.7^{\small\rm o}\,$~\cite{wisa}.  
In Fig.~\ref{plot}  we plot the  $\Xi\pi$-scattering phases as 
a function of the pion momentum.

\begin{center}\begin{minipage}{1\textwidth}  
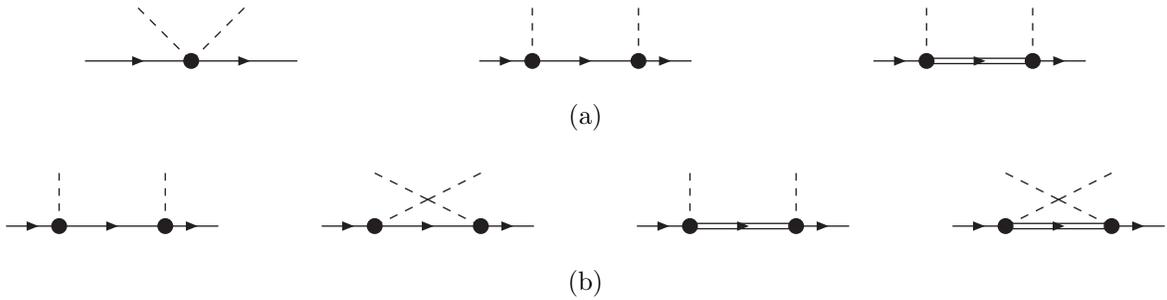
\begin{figure}[t]         
   \hspace*{\fill} 
\begin{picture}(80,50)(-40,-20)    
\ArrowLine(-40,0)(0,0) \DashLine(-20,20)(0,0){3} 
\DashLine(0,0)(20,20){3} \ArrowLine(0,0)(40,0) \Vertex(0,0){3} 
\end{picture}
   \hspace*{\fill} 
\begin{picture}(80,50)(-40,-20)    
\ArrowLine(-40,0)(-20,0) \DashLine(-20,20)(-20,0){3} 
\ArrowLine(-20,0)(20,0) \DashLine(20,0)(20,20){3}   
\ArrowLine(20,0)(40,0) \Vertex(-20,0){3} \Vertex(20,0){3}  
\end{picture}
   \hspace*{\fill} 
\begin{picture}(80,50)(-40,-20)    
\ArrowLine(-40,0)(-20,0) \Line(-20,1)(20,1) \Line(-20,-1)(20,-1) 
\ArrowLine(-1,0)(1,0) \DashLine(-20,20)(-20,0){3} 
\DashLine(20,0)(20,20){3} \ArrowLine(20,0)(40,0) 
\Vertex(-20,0){3} \Vertex(20,0){3}  
\end{picture}  
   \hspace*{\fill} 
\\ 
   \hspace*{\fill} 
\begin{picture}(10,10)(-5,-5)
\Text(0,5)[c]{(a)}   
\end{picture}
   \hspace*{\fill} 
\\       
   \hspace*{\fill} 
\begin{picture}(80,50)(-40,-20)    
\ArrowLine(-40,0)(-20,0) \DashLine(-20,20)(-20,0){3} 
\ArrowLine(-20,0)(20,0) \DashLine(20,0)(20,20){3}   
\ArrowLine(20,0)(40,0) \Vertex(-20,0){3} \Vertex(20,0){3}  
\end{picture}
   \hspace*{\fill} 
\begin{picture}(80,50)(-40,-20)    
\ArrowLine(-40,0)(-20,0) \DashLine(-20,20)(0,10){3} 
\DashLine(0,10)(20,0){3} \ArrowLine(-20,0)(20,0) 
\DashLine(-20,0)(0,10){3}  \DashLine(0,10)(20,20){3}   
\ArrowLine(20,0)(40,0) \Vertex(-20,0){3} \Vertex(20,0){3}  
\end{picture}
   \hspace*{\fill} 
\begin{picture}(80,50)(-40,-20)    
\ArrowLine(-40,0)(-20,0) \Line(-20,1)(20,1) 
\Line(-20,-1)(20,-1) \ArrowLine(-1,0)(1,0) 
\DashLine(-20,20)(-20,0){3} \DashLine(20,0)(20,20){3}   
\ArrowLine(20,0)(40,0) \Vertex(-20,0){3} \Vertex(20,0){3}  
\end{picture}   
   \hspace*{\fill} 
\begin{picture}(80,50)(-40,-20)    
\ArrowLine(-40,0)(-20,0) \DashLine(-20,20)(0,10){3} 
\DashLine(0,10)(20,0){3} \DashLine(-20,0)(0,10){3}  
\DashLine(0,10)(20,20){3} \Line(-20,1)(20,1) \Line(-20,-1)(20,-1) 
\ArrowLine(-1,0)(1,0) \ArrowLine(20,0)(40,0) 
\Vertex(-20,0){3} \Vertex(20,0){3}  
\end{picture}  
   \hspace*{\fill} 
\\ 
   \hspace*{\fill} 
\begin{picture}(10,10)(-5,-5)
\Text(0,5)[c]{(b)}   
\end{picture}
   \hspace*{\fill} 
\\ 
\caption{\label{diagrams}%
Diagrams for  (a) $\,\Xi^0\pi^-\rightarrow\Xi^0\pi^-\,$   
and  (b) $\,\Xi^-\pi^0\rightarrow\Xi^-\pi^0.\,$   
A dashed line denotes a pion field, and a single (double) 
solid-line denotes a $\Xi$ ($\Xi^*$) field.}
\end{figure}             
\end{minipage}\end{center}
\begin{center}\begin{minipage}{1\textwidth}  
\begin{figure}[ht]         
   \hspace*{\fill}   
\begin{picture}(300,220)(-50,-110)   
\LinAxis(0,80)(0,-80)(4,5,3,0,1) \LinAxis(200,80)(200,-80)(4,5,-3,0,1) 
\LinAxis(0,80)(200,80)(4,5,-3,0,1) \LinAxis(0,-80)(200,-80)(4,5,3,0,1)  
\DashLine(145.878,80)(145.878,-80){1} 
\SetWidth{1.0} 
\Curve{(0,0)(2.,0.000246)(4.,0.00198)
(10.,0.0314)(12.,0.0548)(14.,0.088)(16.,0.133)(18.,0.192)
(20.,0.268)(22.,0.364)(24.,0.482)(26.,0.627)(28.,0.802)
(30.,1.01)(32.,1.27)(34.,1.57)(36.,1.93)(38.,2.35)
(40.,2.86)(42.,3.45)(44.,4.16)(46.,5.01)(48.,6.02)
(50.,7.23)(52.,8.69)(54.,10.5)(56.,12.7)(58.,15.4)
(60.,18.9)(62.,23.3)(64.,29.3)(66.,37.5)(68.,49.4)
(70.,67.8)(70.9,79.9)} 
\Curve{(88.,-79.6)(90.,-72.1)(92.,-66.7)(94.,-62.6)(96.,-59.4)(98.,-56.9)
(100.,-55.)(102.,-53.4)(104.,-52.1)(106.,-51.1)(108.,-50.3)
(110.,-49.7)(112.,-49.2)(114.,-48.8)(116.,-48.5)(118.,-48.3)
(120.,-48.2)(122.,-48.1)(124.,-48.1)(126.,-48.2)(128.,-48.2)
(130.,-48.4)(132.,-48.5)(134.,-48.7)(136.,-49.)(138.,-49.2)
(140.,-49.5)(142.,-49.8)(144.,-50.1)(146.,-50.5)(148.,-50.8)
(150.,-51.2)(152.,-51.6)(154.,-52.)(156.,-52.4)(158.,-52.9)
(160.,-53.3)(162.,-53.7)(164.,-54.2)(166.,-54.7)(168.,-55.2)
(170.,-55.6)(172.,-56.1)(174.,-56.6)(176.,-57.1)(178.,-57.7)
(180.,-58.2)(182.,-58.7)(184.,-59.2)(186.,-59.8)(188.,-60.3)
(190.,-60.9)(192.,-61.4)(194.,-62.)(196.,-62.5)(198.,-63.1)(199.,-63.4)}
\DashCurve{(0,0)(5.,0.000405)(10.,0.00322)
(25.,0.0481)(30.,0.0812)(35.,0.126)(40.,0.183)(45.,0.253)
(50.,0.336)(55.,0.433)(60.,0.544)(65.,0.668)(70.,0.807)
(75.,0.958)(80.,1.12)(85.,1.3)(90.,1.49)(95.,1.7)
(100.,1.91)(105.,2.14)(110.,2.38)(115.,2.63)(120.,2.89)
(125.,3.16)(130.,3.44)(135.,3.73)(140.,4.03)(145.,4.34)
(150.,4.66)(155.,4.99)(160.,5.32)(165.,5.66)(170.,6.01)
(175.,6.37)(180.,6.73)(185.,7.1)(190.,7.48)(195.,7.86)
(200.,8.25)}{4}  
\footnotesize  
\rText(-40,0)[][l]{$\delta_{2I}^{}\;(\rm degrees)$}   
\Text(100,-100)[t]{$|\bfm{k}|\;(\rm GeV)$}   
\Text(-5,80)[r]{$20$} \Text(-5,40)[r]{$10$} 
\Text(-5,0)[r]{$0$} \Text(-5,-40)[r]{$-10$} \Text(-5,-80)[r]{$-20$} 
\Text(0,-85)[t]{$0$} \Text(50,-85)[t]{$0.1$} \Text(100,-85)[t]{$0.2$} 
\Text(150,-85)[t]{$0.3$} \Text(200,-85)[t]{$0.4$} 
\end{picture} 
   \hspace*{\fill} 
\\
\caption{\label{plot}%
$\Xi\pi$-scattering phases as a function of the center-of-mass momentum of 
the pion. 
The solid and dashed curves denote  $\delta_1^{}$  and  $\delta_3^{}$, 
respectively.  
The vertical dotted-line marks the momentum in the
$\,\Omega^-\rightarrow\Xi\pi\,$  decay.}  
\end{figure}
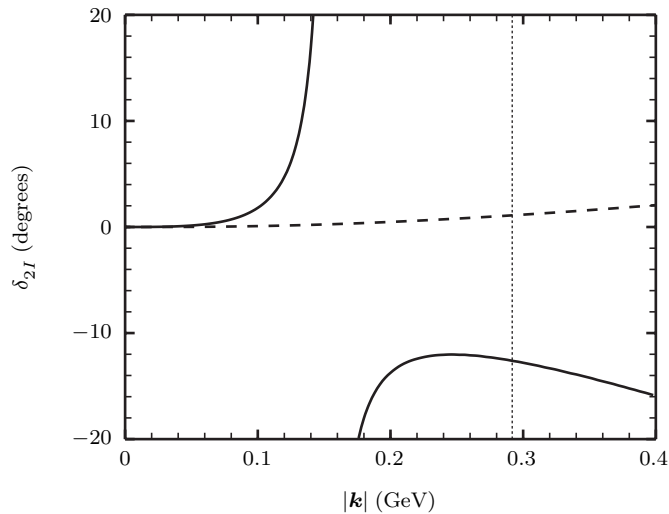             
\end{minipage}\end{center}

As for the weak phases  $\phi_{1,3}^{}$,  within the standard model 
they arise from the $C\!P$-violating phase in the CKM matrix. 
To calculate them, we write the amplitude as  
\begin{equation}   \label{matrix} 
{\rm Re}\,{\cal M}_{\Omega^-\rightarrow\Xi\pi}^{}    
+\ri\, {\rm Im}\,{\cal M}_{\Omega^-\rightarrow\Xi\pi}^{}  \;=\;   
-\langle \Xi\pi| {\cal H}_{\rm eff}^{} |\Omega^- \rangle   \;,  
\end{equation}   
where  ${\cal H}_{\rm eff}^{}$  the short-distance effective 
Hamiltonian that contains a set of four-quark operators 
and describes the $\,|\Delta S| =1\,$  weak interactions 
in the standard model.  
The phases are then given by  
\begin{equation}    
\phi_{1,3}^{}  \;\approx\;  
{ {\rm Im}\,{\cal M}^{(\Omega)}_{1,3}  
 \over  {\rm Re}\,{\cal M}^{(\Omega)}_{1,3} }   \;,  
\end{equation}   
where the subscripts refer to the  $\,|\Delta\bfm{I}|=1/2,3/2\,$  
components of the amplitudes.   
Unfortunately, we cannot compute the matrix elements in~(\ref{matrix})  
in a reliable way. 
In order to estimate them, we employ the vacuum-saturation 
method used in Ref.~\cite{steger}. 
Our calculation yields   
\begin{eqnarray}
\begin{array}{c}   \displaystyle      
\alpha_{3}^{(\Omega)} \re^{{\rm i}\phi_3^{}}  \;=\;  
-0.11  \,+\,  2.8\times10^{-6}\, \ri   \;,
\vspace{2ex} \\    \displaystyle       
\alpha_{1}^{(\Omega)} \re^{{\rm i}\phi_1^{}}  \;=\;  
0.23  \,+\,  2.3\times 10^{-4}\, \ri   \;.
\end{array}
\end{eqnarray}
The  $\,|\Delta\bfm{I}|=3/2\,$  amplitude predicted in vacuum 
saturation is comparable to the one we extract from the data,
$\,\alpha_3^{(\Omega)} = -0.07\pm 0.01.\,$  
To estimate the weak phase, we can obtain the real part of the 
amplitude from experiment and the imaginary part of the amplitude 
from the vacuum-saturation estimate to get 
$\,\phi_3^{}\approx -4\times 10^{-5}.\,$  
Unlike its  $\,|\Delta\bfm{I}|=3/2\,$  counterpart, 
the $\,|\Delta\bfm{I}|=1/2\,$  amplitude is predicted to be about 
a factor of four below the fit.
Taking the same approach as that in estimating  $\phi_3^{}$  results in 
$\,\phi_1^{}\approx 3\times 10^{-4}.\,$  
We can also take the phase directly from the vacuum-saturation estimate 
(assuming that both the real and imaginary parts of the amplitude 
are enhanced in the same way by the physics that is missing from this 
estimate) to find  $\,\phi_1^{} = 0.001.\,$

To summarize, we have  
\begin{eqnarray}
\begin{array}{rcl}   \displaystyle   
{\alpha_{3}^{(\Omega)}\over \alpha_{1}^{(\Omega)}}  &\!\approx&\!  
-0.07   \;,  
\vspace{2ex} \\    \displaystyle       
|\sin \bigl( \delta_3^{}-\delta_1^{} \bigr) |  &\!\approx&\!  0.24   \;,  
\vspace{2ex} \\    \displaystyle       
|\sin(\phi_3^{}-\phi_1)|  &\!\approx&\!  
3\times 10^{-4} \;~{\rm or}~\; 0.001   \;,  
\end{array}   \label{numbers}
\end{eqnarray}  
where the first number for the weak phases corresponds to the 
conservative approach of taking only the imaginary part of the 
amplitudes from the vacuum-saturation estimate and the second 
number is the phase predicted by the model.  
The resulting rate asymmetry is   
\begin{eqnarray}
\bigl| \Delta \bigl( \Xi^0\pi^- \bigr) \bigr| = 
   7\times 10^{-6} \;~{\rm or}~\; 2\times 10^{-5}   \;,  
\end{eqnarray}  
where the difference between these two numbers can be taken as 
a crude measure of the uncertainty in the evaluation of 
the weak phases.  
For comparison, estimates of rate asymmetries in the octet-hyperon 
decays~\cite{donpa} result in values of less than~$\,10^{-6}.$

Using the results of a model-independent study of $C\!P$ violation 
beyond the standard model in octet-hyperon decays in  
Ref.~\cite{heval},   
we can expect that the $C\!P$-violating rate asymmetry 
in  $\,\Omega^-\rightarrow\Xi^0\pi^-\,$  could be ten times larger 
than our estimate above if new physics is responsible for 
$C\!P$ violation.
The upper bound in this case arises from the constraint imposed  on 
new physics by the value of  $\epsilon$  because the P-waves involved 
are parity conserving.

In conclusion, we have estimated the $C\!P$-violating rate 
asymmetry in  $\,\Omega^-\rightarrow\Xi^0\pi^-.\,$  
Within the standard model, it is about  $\,2 \times 10^{-5},\,$  
and it could be up to ten times larger if $C\!P$ violation arises 
from new physics.  
Although there are significant uncertainties in our estimates,  
it is probably safe to say that the rate asymmetry in  
$\,\Omega^-\rightarrow\Xi\pi\,$  decays is much larger 
than the corresponding asymmetries in octet-hyperon decays.

\acknowledgments  

The material presented here has been based on a recent paper 
done in collaboration with G.~Valencia.  
This work  was supported in
part by DOE under contract number DE-FG02-92ER40730.

\end{document}